\documentclass[aps,prd,groupedaddress,showpacs]{revtex4}
\usepackage{amssymb}
\usepackage{amsmath}
\usepackage{amsfonts}
\usepackage{epsfig}
\usepackage{appendix}

\def\qq{\langle \bar q q \rangle}

\newcommand{\seq}{\begin{subequations}}
\newcommand{\sen}{\end{subequations}}
\newcommand{\eq}{\begin{eqnarray}}
\newcommand{\en}{\end{eqnarray}}

\begin{document}

\title{The $g_{\Sigma_Q\Sigma_Q\pi}$ Coupling Constant via Light Cone QCD Sum Rules }
\noindent
\author{
       K. Azizi$^{1*}$,
       M. Bayar$^{2\S}$,
       A. Ozpineci$^{3\dag}$,
       Y. Sarac$^{4\ddag}$
      }\\

\affiliation{\vspace*{1.2\baselineskip}\\$^1$ Physics Division,
Faculty of Arts and
Sciences, Do\u gu\c s University\\
$^{*}$ kazizi@dogus.edu.tr \vspace*{1.2\baselineskip} \\$^2$
Department of Physics, Kocaeli University, 41380 Izmit, Turkey\\
$^{\S}$ melahat.bayar@kocaeli.edu.tr
\vspace*{1.2\baselineskip} \\
$^3$ Physics
Department, Middle East Technical University, 06531, Ankara, Turkey\\
$^{\dag}$ ozpineci@metu.edu.tr\vspace*{1.2\baselineskip} \\
$^4$ Electrical and Electronics Engineering Department, Atilim University, 06836 Ankara, Turkey\\
$^{\ddag}$ ysoymak@atilim.edu.tr \vspace*{1.2\baselineskip} }

\date{\today}

\begin{abstract}
Using the most general form of the interpolating currents,  the
coupling constants $g_{\Sigma_b\Sigma_b\pi}$ and
$g_{\Sigma_c\Sigma_c\pi}$ are calculated within the light cone QCD
sum rules approach. It is found that
$g_{{\Sigma_{c}\Sigma_{c}\pi}}=-8.0\pm1.7$ and
$g_{{\Sigma_{b}\Sigma_{b}\pi}}=-11.0\pm2.1$.

\end{abstract}


\maketitle

\section{Introduction}
Theoretical studies on  heavy baryons containing a single heavy
quark have gained pace recently as a result of the experimental
progresses in the last few years (see \cite{MizukKreps} and
references therein). The presence of  a heavy $b$ or $c$ quark in
these baryons makes them more attractive, theoretically. They can
provide information on the structure of the QCD and its parameters
as well as  clues on new physics effects. For this reason, it is
important to understand their properties as precisely as possible.
Couplings of these baryons to pions are one of the important
properties, also because their pion cloud can significantly modify
their properties.

In light of the experimental developments, the mass spectrum of
these baryons has been studied extensively via QCD sum rules method
both in the finite mass limit (see e.g. \cite{Navarra}) and heavy
quark limit \cite{Shuryak}, in the heavy quark effective
theory\cite{Grozin} and different phenomenological models (see for
example \cite{ Mathur, Ebert, Karliner1, Karliner2, Rosner,
Karliner3}). Electromagnetic
\cite{Julia,Schollamd,Faessler,Patel,An,Aliev1, Aliev2} and strong
\cite{Huang1,Wang4,Azizi,Cheng} interactions are two of other
important characteristic of hadrons that can be used to probe their
structures.

In the present work, the coupling of the $\Sigma_Q$ baryons, $Q=c$
or $b$, to pion is calculated in the light cone QCD sum rules
(LCQSR) framework. This problem has been studied using chiral
perturbation theory in \cite{Cheng}. Note that, the same framework
has also been used to calculate the strong coupling constants of the
light mesons with heavy baryons \cite{Huang1,Wang4,Azizi}. The
layout of the paper is as follows: in the next section, we obtain
sum rules for the corresponding coupling constants in the framework
of
 LCQSR. Section 3 contains our numerical analysis and
discussion.

\section{ Light cone QCD sum rules for the $g_{\Sigma_Q\Sigma_Q\pi}$ coupling constant}

To study the coupling constant in LCQSR method, the following
correlation function is studied:
\begin{equation}\label{T}
\Pi=i\int d^{4}xe^{ipx}\langle \pi(q)\mid {\cal
T}\{\eta_{\Sigma^{ud}_{Q}}(x)\bar{\eta}_{\Sigma^{uu}_{Q}}(0) \}\mid0\rangle,
\end{equation}
where $\eta_{\Sigma^{q_1q_2}_{Q}}(x)$ is the interpolating current
of $\Sigma^{q_1q_2}_{Q}$ baryon containing the quarks $q_1q_2Q$
($q_i=u,~\mbox{or }d$) and ${\cal T}$ stands for the time ordered
product. The correlator can be calculated in terms of hadronic
parameters inserting a complete set of hadronic states. This
expression is  called the phenomenological or physical side. It can
also be calculated in theoretical side in terms of QCD parameters
via operator product expansion (OPE) in deep Euclidean region, where
$-p^2\rightarrow\infty$ and $-(p+q)^2\rightarrow\infty$.  QCD sum
rules for the considered coupling constant is obtained matching both
sides of the correlation function and applying  Borel transformation
in order to suppress contribution of the higher states and
continuum.

First, we focus our attention to calculate the physical side of the
correlation function. The phenomenological side of the sum rules for
the heavy baryons is similar to the light baryons \cite{Aliev3}. For
this aim, we insert a complete set of hadronic state having the same
quantum numbers of the interpolating currents into the correlator.
As a result, we obtain
\begin{eqnarray}\label{T2}
\Pi&=&\frac{\langle0\mid \eta_{\Sigma_{Q}}\mid
\Sigma_{Q}(p_{2})\rangle}{p_{2}^{2}-m_{\Sigma_{Q}}^{2}}\langle
\Sigma_{Q}(p_{2}) \pi(q)\mid \Sigma_{Q}(p_{1})\rangle\frac{\langle
\Sigma_{Q}(p_{1})\mid
\bar{\eta}_{\Sigma_{Q}}\mid 0\rangle}{p_{1}^{2}-m_{\Sigma_{Q}}^{2}}+...,\nonumber\\
\end{eqnarray}
where $p_1=p+q$ and $p_2=p$ and ... denotes the contribution of the
higher states and continuum. The matrix element
creating   baryon from the vacuum is parameterized as:
\begin{equation}\label{lambdabey}
\langle0\mid \eta_{\Sigma_{Q}}\mid
\Sigma_Q(p,s)\rangle=\lambda_{\Sigma_Q}u_{\Sigma_Q}(p,s),
\end{equation}
where $\lambda_{\Sigma_Q}$ is the residue and $u_{\Sigma_Q}$ is the
spinor for the $\Sigma_Q$ baryon. The remaining matrix element
appearing in Eq.~(\ref{T2}) defines the coupling constant
$g_{\Sigma_Q\Sigma_Q\pi}$ through
\begin{eqnarray}\label{matelpar} \langle \Sigma_{Q}(p_{2}){\pi}(q)\mid
\Sigma_{Q}(p_{1})\rangle &=&g_{\Sigma_{Q}\Sigma_{Q}\pi}\overline{u}(p_{2})i\gamma_{5}u(p_{1}).\nonumber\\
\end{eqnarray}
Using the Eqs.~(\ref{lambdabey}) and (\ref{matelpar}) in
Eq.~(\ref{T2}) we obtain the phenomenological side of the correlator
as:
\begin{eqnarray}\label{final phenpart}
\Pi&=&i\frac{g_{\Sigma_{Q}\Sigma_{Q}\pi}|\lambda_{\Sigma_{Q}}|^2}{(p_{1}^{2}-m_{\Sigma_{Q}}^{2})(p_{2}^{2}
-m_{\Sigma_{Q}}^{2})} \left[-\not\!p
\not\!q\gamma_{5}-m_{\Sigma_{Q}}\not\!q\gamma_{5}\right].
\end{eqnarray}
In principle, one can choose any of the structures,
$\not\!p\not\!q\gamma_5$ and $\not\!q\gamma_5$,  existing in Eq.
(\ref{final phenpart}). Our calculations show that  the structure
$\not\!p\not\!q\gamma_5$ leads to a more reliable result.

After obtaining  the physical side,  we proceed to
calculate the QCD side of the correlation function. For this purpose, we use the interpolating currents
in the following general form:
\begin{eqnarray}\label{currentguy}
\eta_{\Sigma^{ud}_{Q}} & = & \frac{1}{\sqrt2}\epsilon^{abc}\Bigg[
\bigg(u_a^T C Q_b \bigg) \gamma^5d_c + \beta \bigg(u_a^T C \gamma^5
Q_b \bigg)d_c -\bigg(Q_a^T C d_b \bigg) \gamma^5u_c - \beta
\bigg(Q_a^T C \gamma^5 d_b \bigg)u_c\Bigg],
\nonumber \\
\eta_{\Sigma^{uu}_{Q}} & = & -\frac{1}{2}\epsilon^{abc}\Bigg[ \bigg(u_a^T
C Q_b \bigg) \gamma^5u_c + \beta \bigg(u_a^T C \gamma^5 Q_b
\bigg)u_c -\bigg(Q_a^T C u_b \bigg) \gamma^5u_c - \beta \bigg(Q_a^T
C \gamma^5 u_b \bigg)u_c\Bigg].
\end{eqnarray}
In Eq.~(\ref{currentguy}) the $Q$ denotes the heavy quarks $b$ or
$c$, $\beta$ is an arbitrary parameter with $\beta=-1$ corresponding
to the Ioffe current, $C$ is the charge conjugation operator and
$a,b$ and $c$ represent the color indices.

To proceed in our  calculations in QCD side,  we need also the
propagators of the heavy and light quarks. They are given as
\cite{Balitsky}:
\begin{eqnarray}\label{heavylightguy}
 S_Q (x)& =&  S_Q^{free} (x) - i g_s \int \frac{d^4 k}{(2\pi)^4}
e^{-ikx} \int_0^1 dv \Bigg[\frac{\not\!k + m_Q}{( m_Q^2-k^2)^2}
G^{\mu\nu}(vx) \sigma_{\mu\nu} + \frac{1}{m_Q^2-k^2} v x_\mu
G^{\mu\nu} \gamma_\nu \Bigg],
\nonumber \\
S_q(x) &=&  S_q^{free} (x)  - \frac{\langle \bar q q \rangle}{12} -
\frac{x^2}{192} m_0^2 \langle \bar q q \rangle  \nonumber \\ &&
 - i g_s \int_0^1 du \left[\frac{\not\!x}{16 \pi^2 x^2} G_{\mu \nu} (ux) \sigma_{\mu \nu} - u x^\mu
G_{\mu \nu} (ux) \gamma^\nu \frac{i}{4 \pi^2 x^2} \right].
 \end{eqnarray}
 The free light and heavy quark propagators in Eq.~(\ref{heavylightguy}) are given in $x$
 representation as
 \begin{eqnarray}\label{free1guy}
S^{free}_{q} &=&\frac{i\not\!x}{2\pi^{2}x^{4}},\nonumber\\
S^{free}_{Q}
&=&\frac{m_{Q}^{2}}{4\pi^{2}}\frac{K_{1}(m_{Q}\sqrt{-x^2})}{\sqrt{-x^2}}-i
\frac{m_{Q}^{2}\not\!x}{4\pi^{2}x^2}K_{2}(m_{Q}\sqrt{-x^2}),\nonumber\\
\end{eqnarray}
where $K_{i}$ are the Bessel functions. Besides the propagators, the
matrix elements of the form $\langle\pi(q)|\bar{q}(x_1)\Gamma_i
q(x_2)|0\rangle$ are also required. Here $\Gamma_i$ represents any
member of the Dirac basis i.e., $\{1, \gamma_{\alpha},
\sigma_{\alpha\beta}/\sqrt{2}, i\gamma_{5}\gamma_{\alpha},
\gamma_{5}\}$. The matrix elements
$\langle\pi(q)|\bar{q}(x_1)\Gamma_i q(x_2)|0\rangle$ are
parameterized in terms of the pion light cone distribution
amplitudes and they are given explicitly as~\cite{R21,R22} \\*
\begin{eqnarray} \langle {\pi}(p)| \bar
q(x) \gamma_\mu \gamma_5 q(0)| 0 \rangle &=& -i f_{\pi} p_\mu
\int_0^1 du  e^{i \bar u p x}
    \left( \varphi_{\pi}(u) + \frac{1}{16} m_{\pi}^2 x^2 {\mathbb A}(u) \right)
\nonumber \\
    &-& \frac{i}{2} f_{\pi} m_{\pi}^2 \frac{x_\mu}{px} \int_0^1 du e^{i \bar u px} {\mathbb B}(u),
\nonumber \\
\langle {\pi}(p)| \bar q(x) i \gamma_5 q(0)| 0 \rangle &=& \mu_{\pi}
\int_0^1 du e^{i \bar u px} \varphi_P(u),
\nonumber \\
\langle {\pi}(p)| \bar q(x) \sigma_{\alpha \beta} \gamma_5 q(0)| 0
\rangle &=& \frac{i}{6} \mu_{\pi} \left( 1 - \tilde \mu_{\pi}^2
\right) \left( p_\alpha x_\beta - p_\beta x_\alpha\right)
    \int_0^1 du e^{i \bar u px} \varphi_\sigma(u),
\nonumber \\
\langle {\pi}(p)| \bar q(x) \sigma_{\mu \nu} \gamma_5 g_s G_{\alpha
\beta}(v x) q(0)| 0 \rangle &=&
    i \mu_{\pi} \left[
        p_\alpha p_\mu \left( g_{\nu \beta} - \frac{1}{px}(p_\nu x_\beta + p_\beta x_\nu) \right)
\right. \nonumber \\
    &-& p_\alpha p_\nu \left( g_{\mu \beta} - \frac{1}{px}(p_\mu x_\beta + p_\beta x_\mu) \right)
\nonumber \\
    &-& p_\beta p_\mu \left( g_{\nu \alpha} - \frac{1}{px}(p_\nu x_\alpha + p_\alpha x_\nu) \right)
\nonumber \\
    &+& p_\beta p_\nu \left. \left( g_{\mu \alpha} - \frac{1}{px}(p_\mu x_\alpha + p_\alpha x_\mu) \right)
        \right]
\nonumber \\
    &\times& \int {\cal D} \alpha e^{i (\alpha_{\bar q} + v \alpha_g) px} {\cal T}(\alpha_i),
\nonumber \\
\langle {\pi}(p)| \bar q(x) \gamma_\mu \gamma_5 g_s G_{\alpha \beta}
(v x) q(0)| 0 \rangle &=&
    p_\mu (p_\alpha x_\beta - p_\beta x_\alpha) \frac{1}{px} f_{\pi} m_{\pi}^2
        \int {\cal D}\alpha e^{i (\alpha_{\bar q} + v \alpha_g) px} {\cal A}_\parallel (\alpha_i)
\nonumber \\
    &+& \left[
        p_\beta \left( g_{\mu \alpha} - \frac{1}{px}(p_\mu x_\alpha + p_\alpha x_\mu) \right) \right.
\nonumber \\
    &-&     p_\alpha \left. \left(g_{\mu \beta}  - \frac{1}{px}(p_\mu x_\beta + p_\beta x_\mu) \right) \right]
    f_{\pi} m_{\pi}^2
\nonumber \\
    &\times& \int {\cal D}\alpha e^{i (\alpha_{\bar q} + v \alpha _g) p x} {\cal A}_\perp(\alpha_i),
\nonumber
\end{eqnarray}
\begin{eqnarray}
\langle {\pi}(p)| \bar q(x) \gamma_\mu i g_s G_{\alpha \beta} (v x)
q(0)| 0 \rangle &=&
    p_\mu (p_\alpha x_\beta - p_\beta x_\alpha) \frac{1}{px} f_{\pi} m_{\pi}^2
        \int {\cal D}\alpha e^{i (\alpha_{\bar q} + v \alpha_g) px} {\cal V}_\parallel (\alpha_i)
\nonumber \\
    &+& \left[
        p_\beta \left( g_{\mu \alpha} - \frac{1}{px}(p_\mu x_\alpha + p_\alpha x_\mu) \right) \right.
\nonumber \\
    &-&     p_\alpha \left. \left(g_{\mu \beta}  - \frac{1}{px}(p_\mu x_\beta + p_\beta x_\mu) \right) \right]
    f_{\pi} m_{\pi}^2
\nonumber \\
    &\times& \int {\cal D}\alpha e^{i (\alpha_{\bar q} + v \alpha _g) p x} {\cal V}_\perp(\alpha_i),
\end{eqnarray}
where $\mu_{\pi} = f_{\pi} \frac{m_{\pi}^2}{m_{u} + m_{d}},$ $\tilde
\mu_{\pi} = \frac{{m_{u} + m_{d}}}{m_{\pi}}$, ${\cal D} \alpha =
 d \alpha_{\bar q}  d \alpha_q  d \alpha_g
\delta(1-\alpha_{\bar q}-\alpha_q-\alpha_g)$ and the
$\varphi_{\pi}(u),$ $\mathbb A(u),$ $\mathbb B(u),$ $\varphi_P(u),$
$\varphi_\sigma(u),$ ${\cal T}(\alpha_i),$ ${\cal
A}_\perp(\alpha_i),$ ${\cal A}_\parallel(\alpha_i),$ ${\cal
V}_\perp(\alpha_i)$ and ${\cal V}_\parallel(\alpha_i)$ are functions
of definite twist and their expressions will be given in the
numerical analysis section. Using these inputs, the QCD side of the
correlator can be calculated in a straightforward fashion.

Having calculated the correlation function both in physical and QCD sides, we equate the
 coefficients of the selected structure from both representations and  apply  Borel
transformation with respect to variables $p^2$ and $(p+q)^2$  to
suppress the contribution of higher states and continuum. As a
result of these  procedures, the QCD sum rules for coupling constant
$g_{\Sigma_Q\Sigma_Q\pi}$ is obtained as:
\begin{eqnarray}\label{magneticmoment2}
e^{\frac{-m_{\Sigma_Q}}{M^2}}m_{\Sigma_Q}|\lambda_{\Sigma_Q}|^2g_{\Sigma_Q\Sigma_Q\pi}&=&
\int_{m_{Q}^{2}}^{s_{0}}e^{-{\frac{s}{M^{2}}}-\frac{m_{\pi}^{2}}{4M^2}}\rho(s)ds+e^{{-\frac{m_Q^2}{M^{2}}}-\frac{m_{\pi}^{2}}{4M^2}}\Gamma,
\end{eqnarray}
where the functions, $\rho(s)$ and $\Gamma$ are given as:
\begin{eqnarray}\label{rho1}
\rho(s)&=& -\frac{1}{64\sqrt{2}\pi^2} \Bigg[2(1-\beta)^{2} m_Q^3
f_{\pi}\{2 \psi_{10}-\psi_{20}+\psi_{31}-2
\ln(\frac{s}{m_{Q}^{2}})\}\varphi_{\pi}(u_{0})\nonumber\\&&-2
(1-\beta^{2})(\psi_{20}-\psi_{31})m_Q^{2}
(-1+\widetilde{\mu_{\pi}}^2) \mu_{\pi} \varphi_{\sigma}(u_{0})+2 m_Q
(1-\beta)^{2} (\psi_{10}-\psi_{21}) A(u_{0}) f_{\pi}
m_{\pi}^{2}\nonumber\\&&+2 (1-\beta)\{(1-\beta)m_Q   f_{\pi}[2
(\psi_{10}-\psi_{21})
(\eta_{3}+\eta_{1}-\eta_{2})-(\eta_{1}-\eta_{2})\ln(\frac{s}{m_{Q}^{2}})]m_{\pi}^{2}\nonumber\\&&+(1+\beta)[m_Q^2
(\eta'_{4}-\eta'_{5})(\psi_{10}-\psi_{20}+\psi_{31}-\ln(\frac{s}{m_{Q}^{2}}))+4(\psi_{10}-\psi_{11}-\psi_{12}+\psi_{21})
u_{0} (\eta'_{4}-2\eta'_{5}) m_{\pi}^{2}] m_{\pi}
\nonumber\\&&-2(-1+\beta) m_Q \ln(\frac{s}{m_{Q}^{2}})[(1-\beta)
(\eta_{1}-2 \eta_{2})f_{\pi} m_{\pi}^{2} +(1+\beta)m_{Q}
(\eta'_{4}-2\eta'_{5}) m_{\pi}]\nonumber\\&&+ 2
(-1+\beta)\{2(-1+\beta) (\psi_{10}+\psi_{21}) m_Q
(\eta_{3}+\eta_{1}-\eta_{2}) f_{\pi}m_{\pi}^{2}+ (1+\beta)
[(\psi_{10}-\psi_{20}+\psi_{31}) m_{Q}^{2}(\eta'_{4}-2\eta'_{5})
\nonumber\\&&+4 (\psi_{10}-\psi_{11}-\psi_{12}+\psi_{21}) u_{0}
(\eta_{4}-2\eta_{5}) m_{\pi}^{2}] m_{\pi}\} \Bigg]+\frac{\langle
\bar{u}u\rangle}{12\sqrt{2}}(-1+\beta^2) \psi_{00} f_{\pi}
\varphi_{\pi}(u_{0}), \nonumber\\&&
\end{eqnarray}
\begin{eqnarray}\label{gamma}
\Gamma&=&-\frac{\langle \bar{u}u\rangle}{288\sqrt{2}} (6 m_Q^2 +1)
(-1+\beta^2)m_{0}^{2}f_{\pi}\varphi_{\pi}(u_{0})-\frac{\langle
\bar{u}u\rangle}{216\sqrt{2}M^{4}}\Bigg[(-1+\widetilde{\mu_{\pi}}^2)m_{\pi}\Big(6
M^{4} m_{Q} (3+2 \beta +6
\beta^{2})\nonumber\\&&-\frac{3}{2}m_{0}^{2} m_Q^3 (3+2 \beta +6
\beta^{2})+M^{2} m_{0}^{2} m_Q (5+4 \beta +5
\beta^{2})\Big)\varphi_{\sigma}(u_{0})\Bigg]\nonumber\\&&-\frac{\langle
\bar{u}u\rangle}{1152\sqrt{2}M^{6}}(1-\beta)^{2} f_{\pi} m_{\pi}^{2}
\Big[24 M^{6} - 6 m_{0}^{2} m_Q^4 +5 M^{2} m_{0}^{2} m_Q^2+24 M^{4}
m_Q^2\Big]A(u_0)\nonumber\\&&+\frac{\langle
\bar{u}u\rangle}{96\sqrt{2}M^{4}} (-1+\beta^2)
(\eta_{3}-\eta_{1})f_{\pi} m_{\pi}^{2} ( m_{0}^{2} m_Q^2-4  M^{4}),
\end{eqnarray}
where
\begin{eqnarray}\label{etalar}
\eta_{j} &=& \int {\cal D}\alpha_i \int_0^1 dv f_{j}(\alpha_i)
\delta(\alpha_{ \bar{q}} +(1- v) \alpha_g -  u_0),
\nonumber \\
\eta'_{j} &=& \int {\cal D}\alpha_i \int_0^1 dv f_{j}(\alpha_i)
\delta'(\alpha_{ \bar{q}} + (1-v) \alpha_g -  u_0),
\nonumber \\
\psi_{nm}&=&\frac{{( {s-m_{Q}}^2 )
}^n}{s^m{(m_{Q}^{2})}^{n-m}},\nonumber \\
\end{eqnarray}
 and  $f_{1}(\alpha_i)={\cal V_{\parallel}}(\alpha_i)$, $f_{2}(\alpha_i)={\cal V_{\perp}}(\alpha_i)$,
 $f_{3}(\alpha_i)={\cal A_{\parallel}}(\alpha_i)$, $f_{4}(\alpha_i)={\cal
 T}(\alpha_i)$, $f_{5}(\alpha_i)=v{\cal T}(\alpha_i)$ are the pion distribution amplitudes. Note that,
 in the above equations, the Borel parameter $M^2$  is defined as $M^{2}=\frac{M_{1}^{2}M_{2}^{2}}{M_{1}^{2}+M_{2}^{2}}$ and
$u_{0}=\frac{M_{1}^{2}}{M_{1}^{2}+M_{2}^{2}}$.  Since the masses of
the initial and final baryons are the same,
 we can set $ M_{1}^{2} = M_{2}^{2} $ and $u_{0} =\frac{1}{2}$.

As it is clear from the expression of the coupling constant, we need
also the residue, $\lambda_{\Sigma_{Q}}$, which is calculated using
mass sum rules \cite{Azizi}:
\begin{eqnarray}\label{residu2}
-\lambda_{\Sigma_{Q}}^{2}e^{-m_{\Sigma_{Q}}^{2}/M^{2}}&=&
\int_{m_{Q}^{2}}^{s_{0}}e^{\frac{-s}{M^{2}}}\rho_{1}(s)ds+e^{\frac{-m_Q^2}{M^{2}}}\Gamma_{1},
\end{eqnarray}
with
\begin{eqnarray}\label{residurho1}
\rho_{1}(s)&=&(<\overline{d}d>+<\overline{u}u>)\frac{(\beta^{2}-1)}{64
\pi^{2}} \Bigg\{\frac{m_{0}^{2}}{4
m_{Q}}(6\psi_{00}-13\psi_{02}-6\psi_{11})\nonumber\\&
+&3m_{Q}(2\psi_{10}-\psi_{11}-\psi_{12}+2\psi_{21})\Bigg\}\nonumber
\end{eqnarray}
\begin{eqnarray}
&+&\frac{ m_{Q}^{4}}{2048 \pi^{4}}
[5+\beta(2+5\beta)][12\psi_{10}-6\psi_{20}
+2\psi_{30}-4\psi_{41}+\psi_{42}-12 ln(\frac{s}{m_{Q}^{2}})],\nonumber\\
\end{eqnarray}
\begin{eqnarray}\label{lamgamma1}
\Gamma_{1}&=&\frac{ (\beta-1)^{2}}{24}<\overline{d}d><\overline{u}u>
\left[\vphantom{\int_0^{x_2}}\right.\frac{m_{Q}^{2}m_{0}^{2}}{2 M^{4}}
+\frac{m_{0}^{2}}{4 M^{2}}-1\Bigg],\nonumber\\
\end{eqnarray}
Note that, since only the square of the residue appears in Eq.
(\ref{residu2}), only the magnitude (not the sign) of
$\lambda_{\Sigma_Q}$ can be determined from mass sum rules. This
indeterminacy is also carried into the coupling constant
calculation, and hence sum rules determine the coupling constant
upto a sign.

\section{Numerical Analysis}
This section is devoted to the numerical analysis of the coupling
constant $g_{{\Sigma_{Q}\Sigma_{Q}\pi}}$. The numerical values of
the required input parameters are given as: $\qq(1~GeV) =
-(0.243)^3~GeV^3$, $m_b = 4.7~GeV$, $m_c = 1.27~GeV$,
$m_{\Sigma_{b}} = 5.805~GeV$ ,$m_{\Sigma_{c}} = 2.455~GeV$,
$m_0^2(1~GeV) = (0.8\pm0.2)~GeV^2$ \cite{Belyaev}, $f_\pi=0.131$
\cite{R21,Belyaev}  and $m_\pi=0.135~GeV$. In order to calculate the
coupling constant, the $\pi$-meson wave functions are needed and
 explicit forms of the these wave functions are represented as \cite{R21, R22}
\begin{eqnarray}
\phi_{\pi}(u) &=& 6 u \bar u \left( 1 + a_1^{\pi} C_1(2 u -1) +
a_2^{\pi} C_2^{3 \over 2}(2 u - 1) \right),
\nonumber \\
{\cal T}(\alpha_i) &=& 360 \eta_3 \alpha_{\bar q} \alpha_q
\alpha_g^2 \left( 1 + w_3 \frac12 (7 \alpha_g-3) \right),
\nonumber \\
\phi_P(u) &=& 1 + \left( 30 \eta_3 - \frac{5}{2} \mu_{\pi}^2 \right)
C_2^{1 \over 2}(2 u - 1)
\nonumber \\
&+& \left( -3 \eta_3 w_3  - \frac{27}{20} \mu_{\pi}^2 -
\frac{81}{10} \mu_{\pi}^2 a_2^{\pi} \right) C_4^{1\over2}(2u-1),
\nonumber \\
\phi_\sigma(u) &=& 6 u \bar u \left[ 1 + \left(5 \eta_3 - \frac12
\eta_3 w_3 - \frac{7}{20}  \mu_{\pi}^2 - \frac{3}{5} \mu_{\pi}^2
a_2^{\pi} \right) C_2^{3\over2}(2u-1) \right],
\nonumber \\
{\cal V}_\parallel(\alpha_i) &=& 120 \alpha_q \alpha_{\bar q}
\alpha_g \left( v_{00} + v_{10} (3 \alpha_g -1) \right),
\nonumber \\
{\cal A}_\parallel(\alpha_i) &=& 120 \alpha_q \alpha_{\bar q}
\alpha_g \left( 0 + a_{10} (\alpha_q - \alpha_{\bar q}) \right),
\nonumber\\
{\cal V}_\perp (\alpha_i) &=& - 30 \alpha_g^2\left[
h_{00}(1-\alpha_g) + h_{01} (\alpha_g(1-\alpha_g)- 6 \alpha_q
\alpha_{\bar q}) +
    h_{10}(\alpha_g(1-\alpha_g) - \frac32 (\alpha_{\bar q}^2+ \alpha_q^2)) \right],
\nonumber\\
{\cal A}_\perp (\alpha_i) &=& 30 \alpha_g^2(\alpha_{\bar q} -
\alpha_q) \left[ h_{00} + h_{01} \alpha_g + \frac12 h_{10}(5
\alpha_g-3) \right],
\nonumber \\
B(u)&=& g_{\pi}(u) - \phi_{\pi}(u),
\nonumber \\
g_{\pi}(u) &=& g_0 C_0^{\frac12}(2 u - 1) + g_2 C_2^{\frac12}(2 u -
1) + g_4 C_4^{\frac12}(2 u - 1),
\nonumber \\
{\mathbb A}(u) &=& 6 u \bar u \left[\frac{16}{15} + \frac{24}{35}
a_2^{\pi}+ 20 \eta_3 + \frac{20}{9} \eta_4 +
    \left( - \frac{1}{15}+ \frac{1}{16}- \frac{7}{27}\eta_3 w_3 - \frac{10}{27} \eta_4 \right) C_2^{3 \over 2}(2 u - 1)
    \right. \nonumber \\
    &+& \left. \left( - \frac{11}{210}a_2^{\pi} - \frac{4}{135} \eta_3w_3 \right)C_4^{3 \over 2}(2 u - 1)\right]
\nonumber \\
&+& \left( -\frac{18}{5} a_2^{\pi} + 21 \eta_4 w_4 \right)\left[ 2
u^3 (10 - 15 u + 6 u^2) \ln u \right. \nonumber\\ &+& \left. 2 \bar
u^3 (10 - 15 \bar u + 6 \bar u ^2) \ln\bar u + u \bar u (2 + 13 u
\bar u) \right] \label{wavefns},
\end{eqnarray}
where $C_n^k(x)$ are the Gegenbauer polynomials,
\begin{eqnarray}
h_{00}&=& v_{00} = - \frac13\eta_4,
\nonumber \\
a_{10} &=& \frac{21}{8} \eta_4 w_4 - \frac{9}{20} a_2^{\pi},
\nonumber \\
v_{10} &=& \frac{21}{8} \eta_4 w_4,
\nonumber \\
h_{01} &=& \frac74  \eta_4 w_4  - \frac{3}{20} a_2^{\pi},
\nonumber \\
h_{10} &=& \frac74 \eta_4 w_4 + \frac{3}{20} a_2^{\pi},
\nonumber \\
g_0 &=& 1,
\nonumber \\
g_2 &=& 1 + \frac{18}{7} a_2^{\pi} + 60 \eta_3  + \frac{20}{3}
\eta_4,
\nonumber \\
g_4 &=&  - \frac{9}{28} a_2^{\pi} - 6 \eta_3 w_3 \label{param0}.
\end{eqnarray}
The constants in the Eqs.~(\ref{wavefns}) and (\ref{param0}) were
calculated at the renormalization scale $\mu=1 ~~GeV^{2}$ using QCD
sum rules \cite{R21, R22, R23} and are given as $a_{1}^{\pi} = 0$,
$a_{2}^{\pi} = 0.44$, $\eta_{3} =0.015$, $\eta_{4}=10$, $w_{3} = -3$
and $ w_{4}= 0.2$.

\begin{figure}[h!]
\begin{center}
\includegraphics[width=12cm]{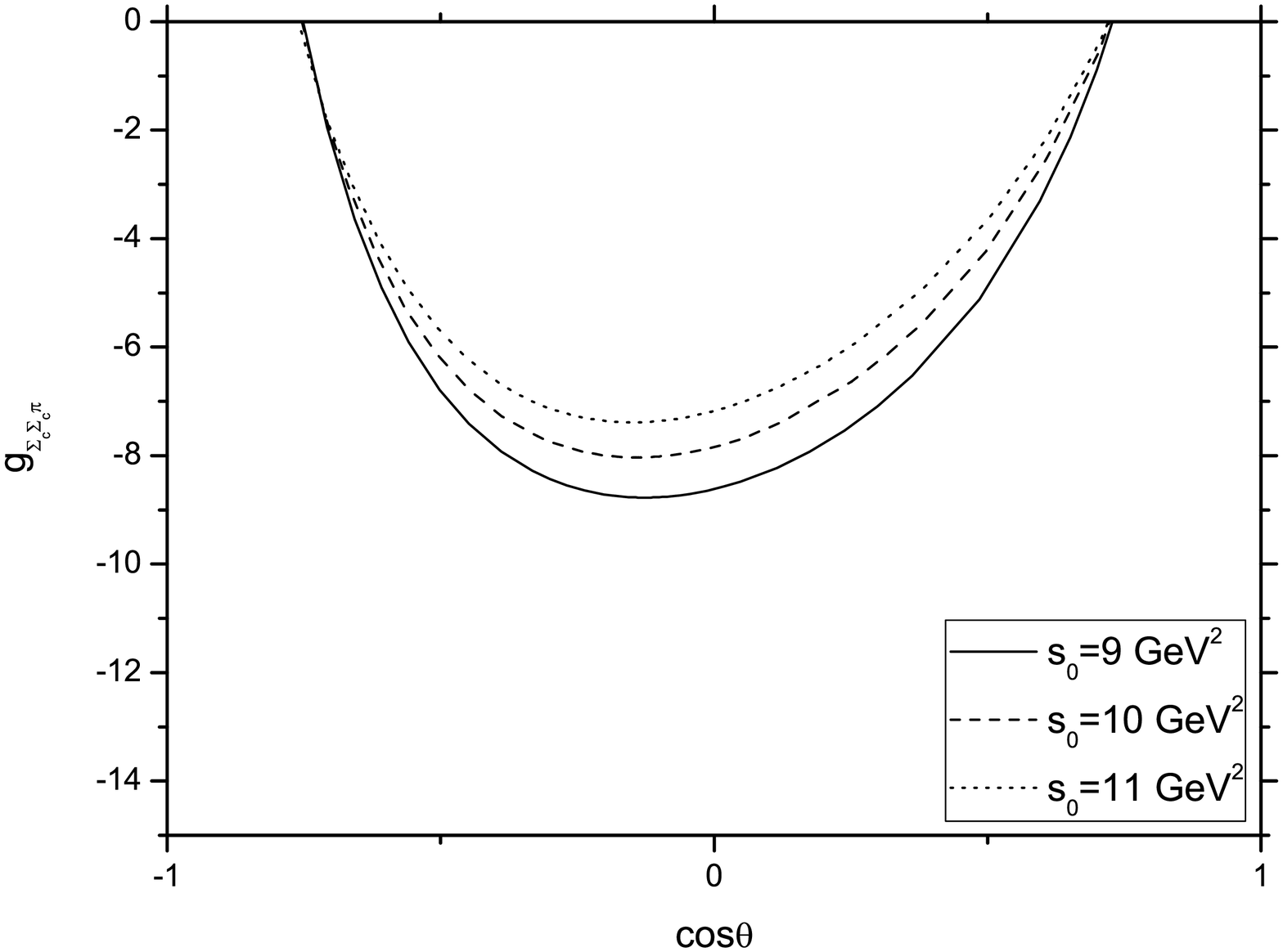}
\end{center}
\caption{The dependence of the coupling constant
$g_{{\Sigma_{c}\Sigma_{c}\pi}}$ on $\cos\theta$ for the value of
Borel parameter $M^{2}=7$ GeV$^2$ and the different values of
continuum threshold $s_{0}$.} \label{fig6}
\end{figure}

\begin{figure}[h!]
\begin{center}
\includegraphics[width=12cm]{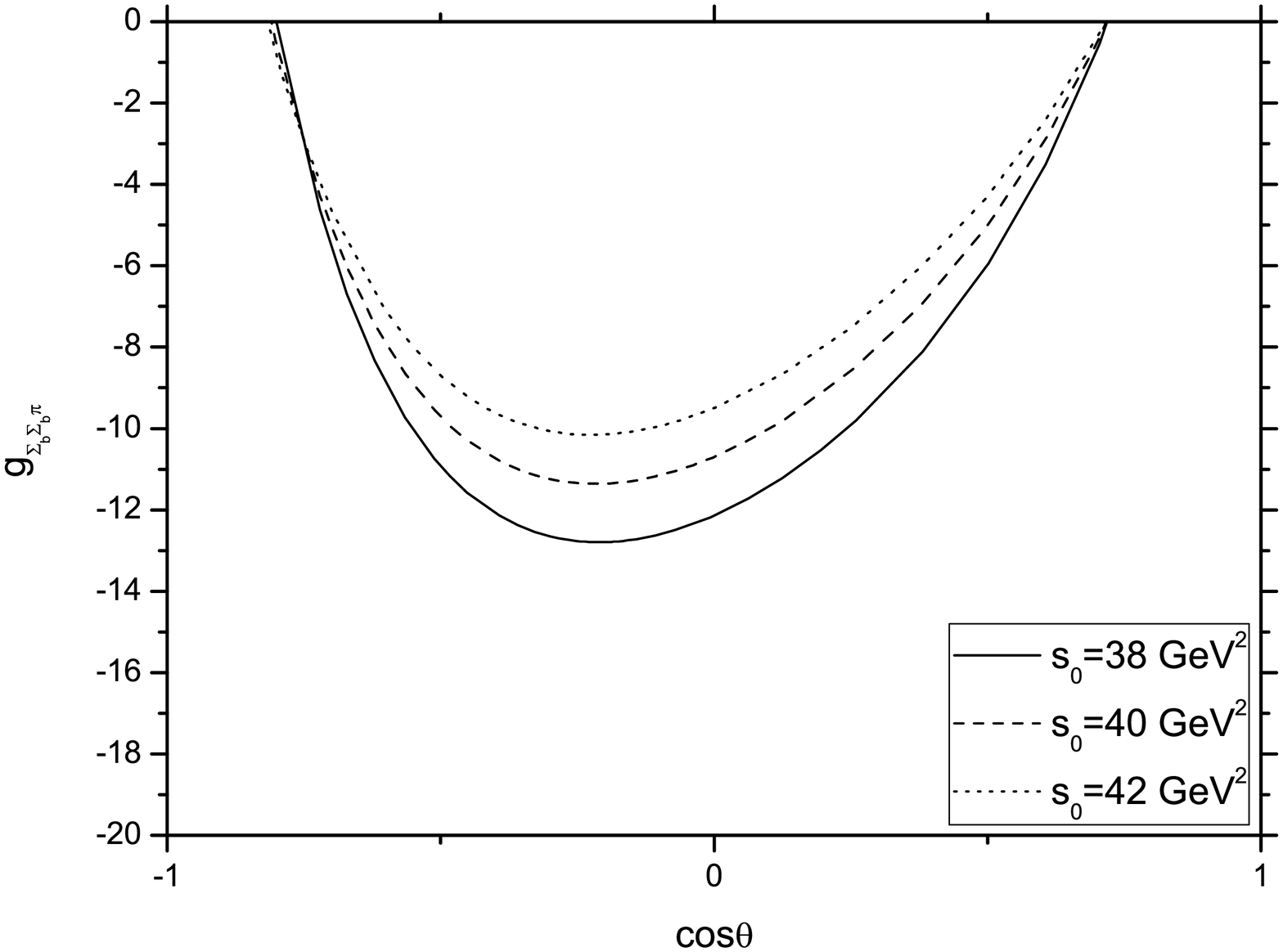}
\end{center}
\caption{The dependence of the coupling constant
$g_{{\Sigma_{b}\Sigma_{b}\pi}}$ on $\cos\theta$ for the value of
Borel parameter $M^{2}=30$ GeV$^2$ and the different values of
continuum threshold $s_{0}$.} \label{fig8}
\end{figure}

\begin{figure}[h!]
\begin{center}
\includegraphics[width=12cm]{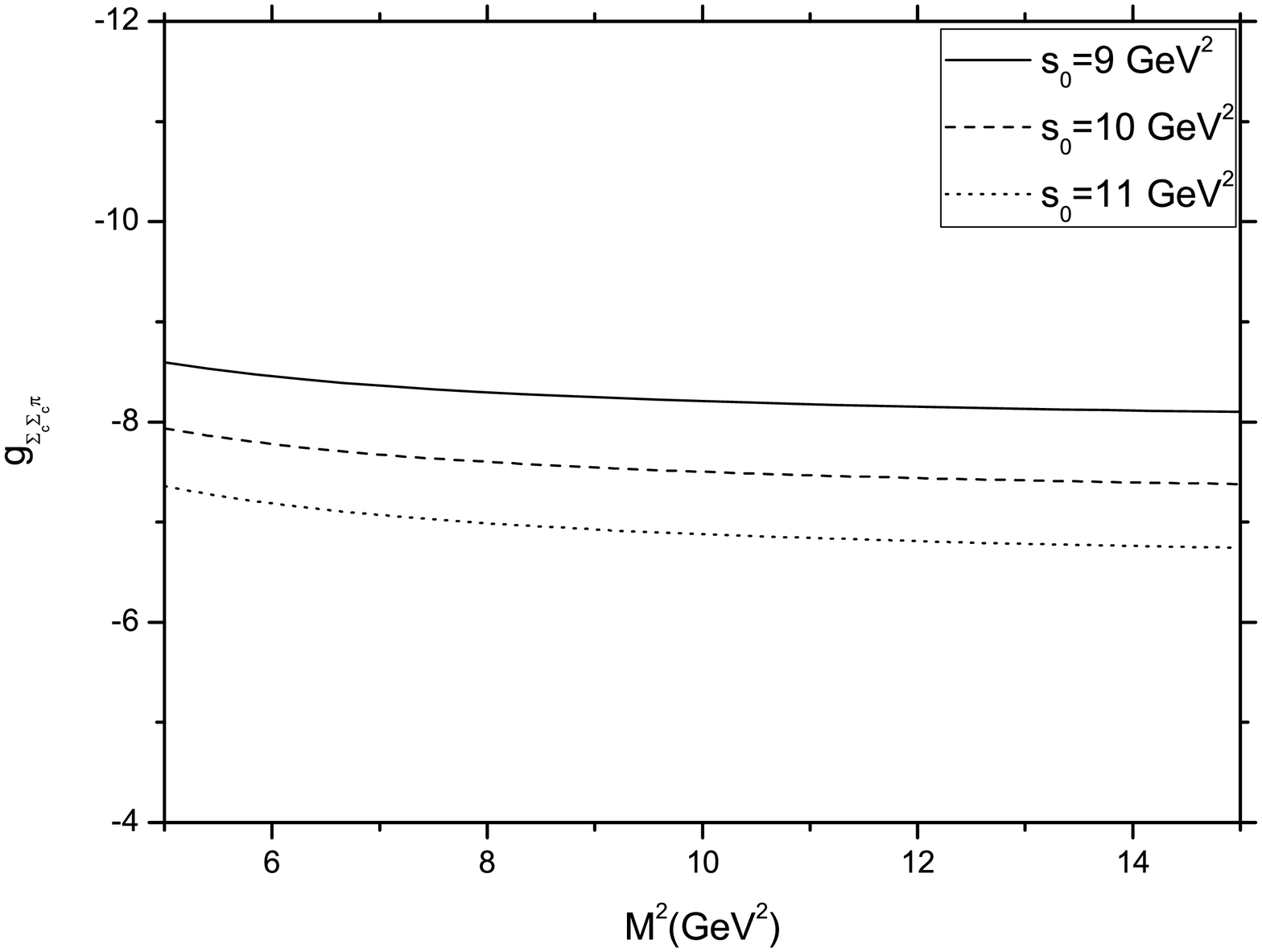}
\end{center}
\caption{The dependence of the coupling constant
$g_{{\Sigma_{c}\Sigma_{c}\pi}}$ on the Borel parameter $M^{2}$ for
the value of arbitrary parameter $\beta=-3$ and the different values
of the continuum threshold $s_{0}$.} \label{fig5}
\end{figure}

\begin{figure}[h!]
\begin{center}
\includegraphics[width=12cm]{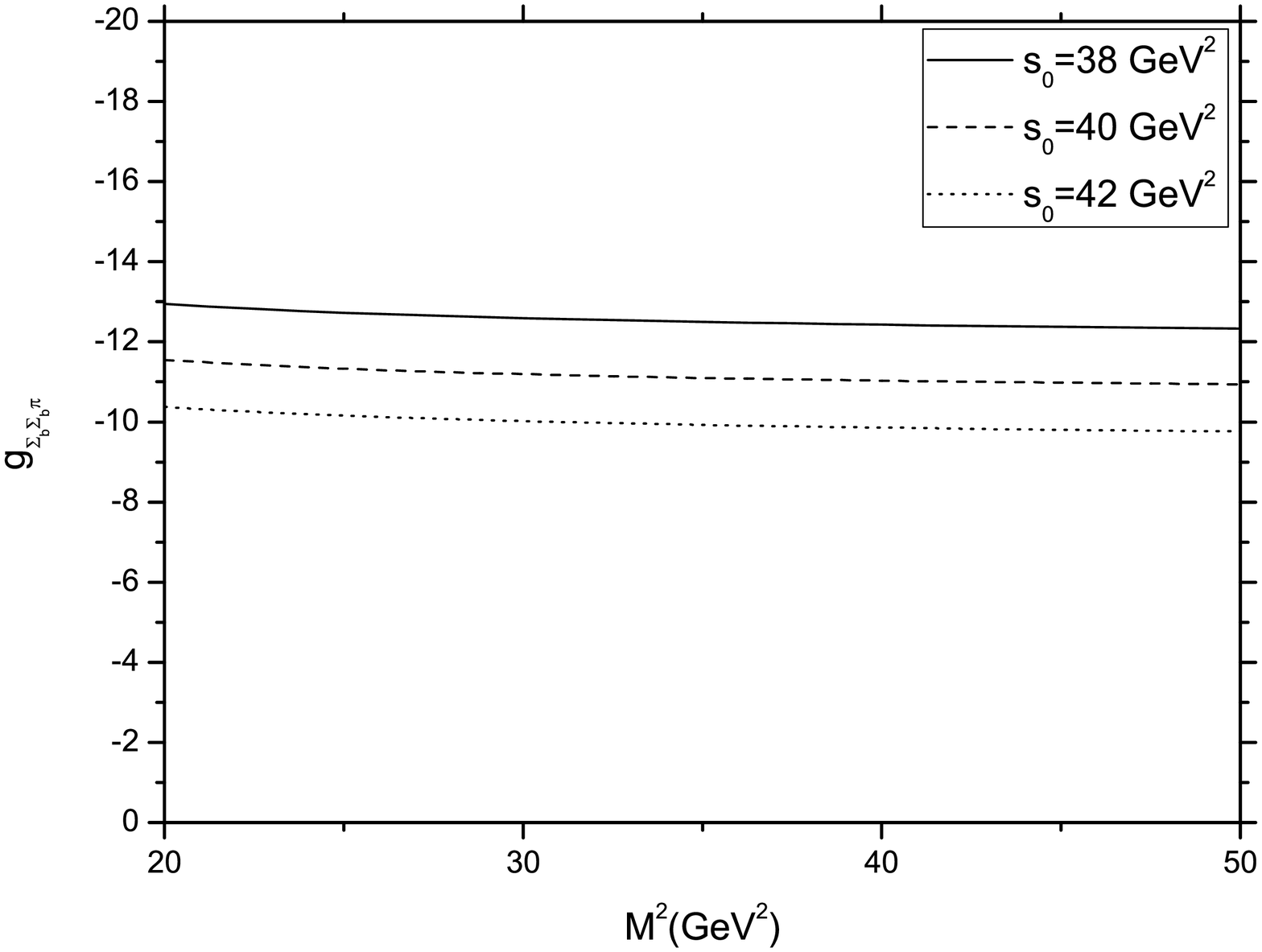}
\end{center}
\caption{The dependence of the coupling constant
$g_{{\Sigma_{b}\Sigma_{b}\pi}}$ on the Borel parameter $M^{2}$ for
the value of arbitrary parameter $\beta=-3$ and the different values
of the continuum threshold $s_{0}$.} \label{fig7}
\end{figure}

The sum rules for the coupling constants contain three auxiliary
parameters, the Borel mass parameter $M^2$, the continuum threshold
$s_0$ and the general parameter $\beta$. These are not physical
parameters, hence the sum rules for the coupling constants should be
independent of them. Therefore, we should look for the working
regions for these parameters such that the dependency on these
parameters are weak. At too large values of the Borel parameter
$M^2$, the suppression of the contribution of higher states and
continuum is reduced, increasing the error due to the quark-hadron
duality approximation. The lower limit of the $M^2$ can be obtained
by requiring that the highest twist contribution, which is
suppressed by a higher power of $M^2$, should be small. Therefore,
the higher states and continuum contributions should comprise a
small percentage of the total dispersion integral. The continuum
threshold $s_{0}$ is not completely arbitrary and is related to the
energy of the first exited state. From the numerical analysis of the
sum rules for  the coupling constants, the continuum thresholds
$s_{0}$ are obtained as $38~GeV^2\leq s_{0}\leq 42~GeV^2$ and
$9~GeV^2 \leq s_{0} \leq 11~GeV^2$ for the $\Sigma_{b}\Sigma_{b}\pi$
and $\Sigma_{c}\Sigma_{c}\pi$ vertexes, respectively. In order to
acquire the working region for $\beta$ parameter, we plot the
$g_{{\Sigma_{Q}\Sigma_{Q}\pi}}$ as a function of $\cos\theta$ in the
interval $-1\leq \cos\theta \leq1$, which corresponds to
$-\infty\leq \beta \leq \infty$, where $\tan\theta=\beta$. From the
Figs. (\ref{fig6}) and  (\ref{fig8}), which show the dependency of
coupling constants on the $\cos\theta$, it is seen that for
$-0.5\leq \cos\theta \leq 0.4$ the dependence of the coupling
constants on $\cos\theta$ is weak. We present also the dependency of
the coupling constants, $g_{{\Sigma_{c}\Sigma_{c}\pi}}$ and
$g_{{\Sigma_{b}\Sigma_{b}\pi}}$, on the Borel parameter $M^{2}$ in
Figs. (\ref{fig5}) and (\ref{fig7}),  respectively. These Figures
reveal that the coupling constants are quite stable in the chosen
Borel mass region. From these Figures, the numerical values of the
coupling constants, $g_{{\Sigma_{c}\Sigma_{c}\pi}}$ and
$g_{{\Sigma_{b}\Sigma_{b}\pi}}$, can be extracted as follows:

\begin{eqnarray}
g_{{\Sigma_{c}\Sigma_{c}\pi}}=-8.0\pm1.7 ,~~~~~~~~~~~~~~~
g_{{\Sigma_{b}\Sigma_{b}\pi}}=-11.0\pm2.1
 \label{couplings}.
\end{eqnarray}
The errors appearing in our predictions are due to the uncertainties
in the input as well as the auxiliary parameters.

At the end of this section, let us compare our results with the
existing prediction on the  $g_{{\Sigma^{+}\Sigma^{0}\pi^{+}}}$
coupling constant \cite{Aliev3} for the light $\Sigma$ baryon. In
light case the $s$ quark   and in our case (baryons with a single
heavy quark) the $c$ or $b$ quark is the spectator quark. These
quarks do not participate in the considered strong interactions, so
one expects these coupling constants (heavy and light cases) have
values close to each other. In \cite{Aliev3}, it was estimated that,
$g_{\Sigma^+\Sigma^0\pi^+}=-9\pm2$. The same transition for the
light $\Sigma$ baryon was also studied in \cite{GErkol} and
\cite{HKim}. In \cite{GErkol}, the ratio,
$g_{\Sigma^+\Sigma^0\pi^+}/g_{NN\pi}\simeq 0.8$ was obtained, which
leads to the value of the coupling constant,
$g_{\Sigma^+\Sigma^0\pi^+}\simeq11.9$ with the experimental value of
$g_{NN\pi}=14.9$ \cite{Arndt}. The
 value of the same coupling constant was  given in \cite{HKim} as  $g_{\Sigma^+\Sigma^0\pi^+}=-11.9\pm0.4$.
 Comparison of our
results with that of \cite{Aliev3,GErkol,HKim} supports the
expectation that the spectator quark does not effect the pionic
coupling, significantly.

\section{Acknowledgment}
The work has been supported in part by the European Union
(HadronPhysics2 project "Study of strongly interacting matter").

\newpage

\end{document}